\documentclass[sigconf,screen,nonacm]{acmart}

\usepackage{booktabs}
\usepackage{graphicx}
\usepackage{xspace}
\usepackage{enumitem}
\usepackage{array}
\usepackage{tabularx}
\usepackage{seqsplit}
\usepackage{url}

\newcommand{\approach}{\textsc{SPIDER4TianoCore}\xspace}
\newcommand{\edkii}{EDK~II\xspace}

\begin{document}

\title{SPIDER4TianoCore: Enhancing Patch-Propagation for the TianoCore UEFI Firmware Development Ecosystem}

\author{Laura Baird}
\email{lbaird@uccs.edu}
\affiliation{%
  \institution{Department of Computer Science, University of Colorado Colorado Springs}
  \state{Colorado}
  \country{USA}
}

\author{Devin Haggitt}
\email{dhaggitt@uccs.edu}
\affiliation{%
  \institution{Department of Computer Science, University of Colorado Colorado Springs}
  \state{Colorado}
  \country{USA}
}

\author{Terrance E. Boult}
\email{tboult@uccs.edu}
\affiliation{%
  \institution{Department of Computer Science, University of Colorado Colorado Springs}
  \state{Colorado}
  \country{USA}
}

\author{Aravind Machiry}
\email{amachiry@purdue.edu}
\affiliation{%
  \institution{School of Electrical and Computer Engineering, Purdue University}
  \state{Indiana}
  \country{USA}
}

\author{Armin Moin}
\email{moin@purdue.edu}
\affiliation{%
  \institution{School of Applied and Creative Computing, Purdue University}
  \state{Indiana}
  \country{USA}
}

\renewcommand{\shortauthors}{Baird et al.}
\renewcommand{\shorttitle}{SPIDER4TianoCore}




\begin{abstract}
We propose and demonstrate \approach, a packaged Python command-line tool that provides integration-stage patch-status evidence for the TianoCore/UEFI firmware supply chain. Given an upstream pre-patch and post-patch pair and prepared downstream targets, the tool reports \textsc{Vulnerable}, \textsc{Already Patched}, \textsc{Not Applicable}, or \textsc{Uncertain} with supporting evidence for maintainer review. Our work is inspired by SPIDER's patch-propagation framing, but \approach does not itself prove that a patch is safe to propagate. We evaluate the engine on 20 prepared target/CVE pairs from eight public downstream \edkii repositories and two CVEs. The analyzers produce 10 high-confidence pre-patch matches and four high-confidence post-patch matches, conservatively abstain on six targets, and make no confidently wrong classifications relative to the recorded manual patch-state labels. These preliminary results demonstrate reproducible evidence generation for prepared targets rather than general downstream accuracy.
\end{abstract}

\maketitle

\section{Introduction}
\label{sec:introduction}

Open-source firmware ecosystems rely on timely security patch propagation. When a vulnerability is fixed upstream, downstream release branches, platform packages, and vendor-maintained forks may remain vulnerable until maintainers identify, adapt, and apply the same fix. This problem is well studied in Linux and related software repositories, where security patches may propagate slowly across forks and stable versions~\cite{Machiry+2020,Shariffdeen+2021a,Li+2024}. It is especially important for firmware, where vulnerable code can execute before the operating system and where mitigation may depend on vendor update pipelines.

TianoCore \edkii is a widely used open-source UEFI implementation. Its security fixes can appear on mainline, release branches, downstream trees, and vendor-derived code bases. This paper addresses a narrow integration-stage research question through the demonstrated tool: After an upstream security fix and candidate target references are known, can a manifest-driven workflow produce reviewable propagation evidence for known \edkii CVEs and candidate downstream references?

\begin{figure}[t]
    \centering
    \includegraphics[width=\linewidth]{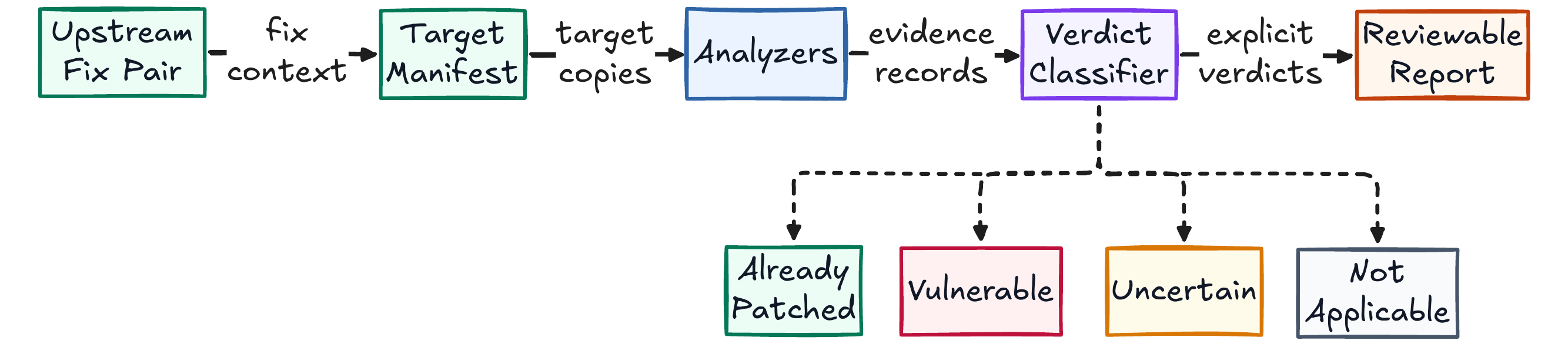}
    \caption{The overall pipeline in \approach. A manifest identifies an upstream pre/post patch pair and target copies. Python-native analyzers produce evidence for a verdict classifier; SPIDER-style condition-equivalence reasoning remains an optional analyzer for future cases where simple evidence is insufficient.}
    \Description{A left-to-right flow from an upstream pre-patch and post-patch pair through a target resolver and analyzers to a verdict classifier and generated report.}
    \label{fig:spidercore-pipeline}
\end{figure}

Our work builds on prior work in the literature, called SPIDER \cite{Machiry+2020}, which introduced the idea of identifying safe patches that restrict the input space of a program without changing valid-output behavior, enabling faster propagation across related repositories~\cite{Machiry+2020}. SPIDER's conceptual framing is a strong fit for firmware maintenance: If a security fix is safe and a downstream copy still contains the vulnerable logic, maintainers need evidence that the downstream copy likely requires attention. However, our integration attempt for using the original SPIDER code with the TianoCore \edkii revealed some challenges concerning the legacy Java code, the Ant tool, the native Z3 Satisfiability Modulo Theories (SMT) solver dependencies, and other cross-platform reproducibility issues. Our tool is not a substitute for SPIDER. Instead, we adapt the propagation-evidence workflow to classify the patch status of prepared downstream targets using Python-native analyzers. This classification does not establish that a patch is safe to propagate, and SPIDER-style condition-equivalence analysis remains outside the current tool. Figure \ref{fig:spidercore-pipeline} illustrates the overall architecture of the pipeline deployed in \approach.

The contribution of this paper is twofold. First, it demonstrates the \approach tool, with an \edkii-specific propagation-evidence workflow that maps an upstream pre/post fix pair and prepared target copies to explicit verdicts, namely \textsc{Vulnerable}, \textsc{Already Patched}, \textsc{Not Applicable}, or \textsc{Uncertain}. \approach is packaged as a Python command-line tool that combines text normalization, tree-sitter-based C analysis, a stable report schema, and public-safe advisory outputs. Second, we publish a provenance-checked benchmark over two public \edkii CVEs and 20 prepared target/CVE pairs from eight public downstream repositories.

The remainder of this paper is structured as follows. Section \ref{sec:related-work} reviews related work in the literature. In Section \ref{sec:approach}, we demonstrate the \approach tool. Section \ref{sec:evaluation} reports the preliminary evaluation results. Finally, we conclude and present future work in Section \ref{sec:conclusion}.

\section{Related Work}
\label{sec:related-work}

\paragraph{Patch Propagation in Firmware Ecosystems}
Patch propagation asks whether a fix introduced in one code base should also be applied to related code bases. Firmware makes this question difficult because release branches, platform packages, vendor forks, and security disclosure constraints can separate the upstream fix from downstream integration. TianoCore \edkii also contains public release-branch backports of security fixes. These backports are useful for early evaluation because they provide real examples of the same fix appearing as different commits in closely related code.

\paragraph{SPIDER and Safe Patches}
SPIDER analyzes source-code changes to identify safe patches that can be propagated to related repositories with limited testing effort~\cite{Machiry+2020,Luo+2024}. It uses fine-grained differencing, program representations such as Abstract Syntax Tree (AST) and Control Flow Graph (CFG) structures, and symbolic reasoning to determine whether a patch restricts inputs while preserving behavior for valid inputs. \approach uses SPIDER's safe-patch and propagation framing, but the current evaluation does not run SPIDER's full symbolic pipeline. Instead, the implemented analyzers are deliberately simple: normalized text matching and identifier-normalized tree-sitter matching over prepared C function inputs. SPIDER/Z3-style condition-equivalence analysis is left as a future escalation path for cases where current analyzers produce an \textsc{Uncertain} result that blocks a real maintainer decision.

\paragraph{Backporting, Transplantation, and Semantic Patching}
Automated backporting and transplantation systems adapt patches to older or related versions. FixMorph synthesizes transformations from Linux mainline patches and backports them to stable versions~\cite{Shariffdeen+2021a}; TSBPORT uses semantic information to improve OSS patch backporting~\cite{Yang+2023}; and PatchWeave studies automated patch transplantation across similar programs~\cite{Shariffdeen+2021b}. Coccinelle and SmPL support semantic patching for C code and have been used for collateral evolution and Linux driver backporting~\cite{Padioleau+2008,RodriguezLawall2015,Thung+2016}. Recent large language model (LLM)-assisted work, including PortGPT, explores patch porting and backporting workflows~\cite{Pan+2024,Li+2026}. These systems are complementary to \approach: after a target is classified as likely vulnerable, a later workflow could adapt or propose a patch. This paper focuses only on conservative propagation-status evidence.

\section{Approach}
\label{sec:approach}

\paragraph{Inputs, Outputs, and Verdicts}
A manifest is a short declarative specification --- in the current implementation, a YAML file --- that names an upstream pre-patch file, an upstream post-patch file, the function of interest, and one or more target copies pinned to specific commits (for example, \texttt{kraxel/edk2@7c0ad2c3}). \approach takes a manifest as input and produces a JSON report, a technical Markdown report, and optionally an advisory-only Markdown artifact: a maintainer-facing summary restricted to verdict and evidence statements, distinct from the public-safe rendering mode described in Section~3.2, which redacts local paths and identifiers before any artifact leaves the development environment. Each target receives one of four verdicts:

\begin{description}[style=nextline,leftmargin=1.6em,labelwidth=1.4em,labelsep=0.4em]
    \item[\textsc{Vulnerable}] The target matches the upstream pre-patch function body under the current analyzer evidence.
    \item[\textsc{Already Patched}] The target matches the upstream post-patch function body under the current analyzer evidence.
    \item[\textsc{Not Applicable}] The affected function/path is absent under the current target specification. This is not proof that the whole repository, other refs, or other packages are unaffected.
    \item[\textsc{Uncertain}] The target is present but too different for the current analyzers to classify safely, or analyzers disagree in a way that requires manual review.
\end{description}

The \textsc{Uncertain} label is a first-class security-maintenance outcome, not a failed run. It prevents the tool from turning weak evidence into an unsupported recommendation. Confidence labels are also tool-side labels: \emph{high} means an analyzer found an exact normalized-text or identifier-normalized tree-sitter match to the upstream pre/post function, while \emph{low} means the current analyzers did not find such a match. Confidence is not exploitability confidence, severity confidence, or release-obligation confidence.

\paragraph{Components and Artifact Boundaries}
The implementation has four components, each with a narrow responsibility: resolving declared targets to concrete files, analyzing them against the upstream fix, classifying a verdict, and aggregating a report. \textbf{TargetResolver} maps manifest entries to local prepared files. A future target-preparation layer may clone or fetch external repositories at pinned revisions, but discovery/ref selection is intentionally separate from analyzer runtime. \textbf{Analyzers} compare each target against upstream pre/post code using normalized text and tree-sitter C function normalization. \textbf{VerdictClassifier} chooses a high-confidence vulnerable or patched match when available, reports \textsc{Not Applicable} only when analyzers agree the function is absent, and otherwise emits \textsc{Uncertain}. \textbf{ReportAggregator} emits a stable schema and human-readable outputs for review.

The current packaged command-line interface (CLI) tool includes typed manifest and report models, a JSON report schema, technical Markdown output, advisory Markdown output, a public-safe rendering mode that suppresses local paths and fixture scaffolding, and a regression suite for the controlled and public downstream-candidate cases. A run-all harness executes all eight manifests, writes internal and public-safe artifacts, measures analyzer compute separately from CLI wall-clock, verifies public-safe nulling, and scans generated artifacts for local-path or identifier leaks. The tool does not submit patches, post GitHub comments, notify vendors, decide embargo handling, or claim patch acceptance. Its defensible claim is narrower: it reduces the time needed to classify prepared targets and package propagation evidence after candidate refs have already been selected.

\section{Preliminary Evaluation}
\label{sec:evaluation}

\paragraph{Research Question and Dataset}
The preliminary evaluation asks whether a manifest-driven workflow can produce conservative, reviewable propagation evidence for known \edkii CVEs and candidate downstream refs, including cases where simple analyzers should refuse to classify. We selected two public \edkii CVE fixes. The first is CVE-2024-38797, an out-of-bounds read fix in \texttt{HashPeImageByType}. The second is CVE-2023-45234, part of the PixieFail class of NetworkPkg Dynamic Host Configuration Protocol version 6 (DHCPv6) client vulnerabilities, represented by \texttt{PxeBcHandleDhcp6Offer}~\cite{Quarkslab2024,TianoCore2026}.

The expanded benchmark contains 20 prepared target/CVE pairs from eight public downstream repositories:
\nolinkurl{microsoft/mu_basecore},
\nolinkurl{linuxboot/edk2},
\nolinkurl{worproject/edk2},
\nolinkurl{Dasharo/edk2},
\nolinkurl{timberland-sig/edk2},
\nolinkurl{system76/edk2},
\nolinkurl{acidanthera/audk},
and \nolinkurl{andreiw/MacchiatoBin-edk2}.
Eleven targets exercise CVE-2023-45234 and nine exercise CVE-2024-38797.
Every prepared file is pinned to a repository URL, commit, and source path and was byte-verified against the recorded source blob.
All 20 target/CVE pairs have recorded manual semantic-review labels; the prepared benchmark, provenance records, and deterministic results are archived in Harvard Dataverse~\cite{Baird+2026}.

\paragraph{Experimental Results}
\begin{table*}[t]
\centering
\small
\setlength{\tabcolsep}{4pt}
\renewcommand{\arraystretch}{1.08}
\caption{Results for 20 authentic target/CVE pairs from eight public downstream repositories (11 PixieFail; nine HashPe). The temporal row is a three-snapshot subset.}
\label{tab:results}
\begin{tabularx}{\textwidth}{
  @{}
  >{\raggedright\arraybackslash}p{0.18\textwidth}
  c
  >{\raggedright\arraybackslash}p{0.17\textwidth}
  >{\raggedright\arraybackslash}p{0.22\textwidth}
  >{\raggedright\arraybackslash}X
  @{}
}
\toprule
Benchmark stratum
  & $N$
  & Reviewed state
  & Tool outcome
  & Interpretation \\
\midrule
Exact/AST pre-match
  & 10
  & Pre-patch
  & 10 \textsc{Vulnerable}
  & 10/10 recognized \\

Exact/AST post-match
  & 4
  & Patched-equivalent
  & 4 \textsc{Already Patched}
  & 4/4 recognized \\

Divergent pre-patch
  & 4
  & Pre-patch
  & 4 \textsc{Uncertain}
  & Conservative abstention \\

Divergent patched
  & 2
  & Patched-equivalent
  & 2 \textsc{Uncertain}
  & 0/2 recognized; manual review required \\
\midrule

All scored targets
  & 20
  & Mixed
  & 10 \textsc{Vuln.}, 4 \textsc{Patched}, 6 \textsc{Uncertain}
  & 6/20 abstentions; 0 confidently wrong \\

Temporal backport subset
  & 3 snapshots
  & Pre/fix/head
  & \textsc{Vuln.} $\rightarrow$ \textsc{Patched} $\rightarrow$ \textsc{Patched}
  & One backport event correctly ordered \\
\bottomrule
\end{tabularx}
\end{table*}

Table~\ref{tab:results} summarizes the results for the 20-target benchmark. All 20 target/CVE pairs received a primary manual semantic review, and a second review of 11 selected divergent or temporal-boundary cases agreed on every overlapping label. \approach recognized all 10 exact or AST-equivalent pre-patch cases and all four exact or AST-equivalent post-patch cases. These recognition rates are mechanical characterization results rather than independent evidence of generalization because the stratum assignment shares parsing and normalization logic with the classifier. The four divergent pre-patch targets and two divergent patched-equivalent targets were conservatively classified as \textsc{Uncertain}, yielding 6/20 abstentions and no confidently wrong classifications relative to the recorded manual labels. The tool also correctly ordered one independent-backport sequence across its pre-fix, fix, and head snapshots; these are three correlated observations of one event, not three independent backports. Finally, the cohort is purposive and contains only eight unique normalized function contents across eight repositories, so the reported rates characterize the prepared benchmark rather than population-level downstream accuracy.

\section{Conclusion and Future Work}
\label{sec:conclusion}
In this paper, we proposed and demonstrated \approach, an integration-stage patch-status evidence tool inspired by SPIDER~\cite{Machiry+2020}. On 20 prepared target/CVE pairs from eight public downstream repositories, the implementation produced 14 high-confidence pre-patch or post-patch matches, conservatively abstained on six targets, and made no confidently wrong classifications relative to the recorded manual labels. The tool does not establish that a patch is safe to propagate and is not a complete patch-propagation or notification system. Future work will compare the deterministic analyzers with an LLM baseline on the same benchmark and investigate a human-approved workflow that prepares a validated code-review branch or draft pull request without autonomous merging.

\section*{Software and Data Availability}
The research prototype and deterministic replay instructions are publicly available at~\cite{TianoShield2026}; the prepared benchmark, provenance records, and results are archived in Harvard Dataverse~\cite{Baird+2026}.

\begin{acks}
This material is based upon work supported by the U.S. National Science Foundation (NSF) under Grant No. 2534021. Any opinions, findings, conclusions, or recommendations expressed in this material are those of the authors and do not necessarily reflect the views of the NSF. In preparing this work, we used generative AI models and tools, including OpenAI GPT and Anthropic Claude models, to assist in generating and revising code and text.

This work is accepted to be presented in the FTA 2026 workshop but is not published in the proceedings, according to the ACM SIGSOFT policy (\url{https://www2.sigsoft.org/policies/pcpolicy/}) that does not allow the work of organizers to be published in the workshop proceedings.
\end{acks}

\bibliographystyle{ACM-Reference-Format}
\bibliography{references}

\end{document}